\documentclass[notitlepage,twocolumn,tighten]{aastex61_custom}

\usepackage{fancyapj}
\usepackage{amsmath}
\usepackage{xspace}
\usepackage{multirow}

\newcommand{\br}[1]{\ensuremath{\left[ #1 \right]} }
\newcommand{\rbr}[1]{\ensuremath{\left( #1 \right)} }

\newcommand{\abs}[1]{\ensuremath{\left| #1 \right|} }
\newcommand{\avg}[1]{\ensuremath{\left< #1 \right>} }

\newcommand{\xmm}{\textit{XMM-Newton}\xspace}
\newcommand{\rxte}{\textit{RXTE}\xspace}

\begin{document}

\title{%
Quasi-periodic pulse amplitude modulation in the accreting millisecond pulsar IGR J00291+5934
}

\author{
    Peter Bult
} 
\affil{
Astrophysics Science Division,
NASA Goddard Space Flight Center,
Greenbelt, MD 20771, USA
}
\author{
    Marieke van Doesburgh
}
\affil{
Anton Pannekoek Institute, University of Amsterdam,
Postbus 94249, 1090 GE Amsterdam, The Netherlands
}
\author{
    Michiel van der Klis
}
\affil{
Anton Pannekoek Institute, University of Amsterdam,
Postbus 94249, 1090 GE Amsterdam, The Netherlands
}

\begin{abstract}
    We introduce a new method for analysing the aperiodic variability
    of coherent pulsations in accreting millisecond X-ray pulsars.
    Our method involves applying a complex frequency correction to the
    time-domain light curve, allowing for the aperiodic modulation of
    the pulse amplitude to be robustly extracted in the frequency domain. 
    We discuss the statistical properties of the resulting modulation spectrum
    and show how it can be correlated with the non-pulsed 
    emission to determine if the periodic and aperiodic variability
    are coupled processes.
    Using this method, we study the 598.88 Hz coherent pulsations of the accreting
    millisecond X-ray pulsar IGR J00291+5934 as observed with the
    \textit{Rossi X-ray Timing Explorer} and \textit{XMM-Newton}.
    We demonstrate that our method easily confirms the known coupling
    between the pulsations and a strong 8 mHz QPO in \textit{XMM-Newton}
    observations. Applying our method to the \textit{RXTE} observations, 
    we further show, for the first time, that the much weaker $20$ mHz
    QPO and its harmonic are also coupled the pulsations.
    We discuss the implications of this coupling and indicate how it
    may be used to extract new information on the underlying accretion
    process.
\end{abstract}

\keywords{
	pulsars: general -- 
	stars: neutron --
	X-rays: binaries --	
	X-rays: individual (IGR J00291+5934)
}

\section{Introduction} 

Accreting millisecond X-ray pulsars (AMXPs) are a subclass of
transient low-mass X-ray binaries (LMXBs) for which the stellar
rotation rate of the central neutron star is directly apparent in the
form of coherent pulsation.  Such pulsations manifest as nearly
sinusoidal oscillations in the X-ray flux with fractional sinusoidal
amplitudes of a few to about ten percent \citep{Patruno2012b}.

Pulsations are thought to be due to magnetically channeled accretion,
such that the impact of the accretion column gives rise to local
emission regions, hotspots, near the stellar magnetic poles.  Due to
the stellar rotation such hotspots undergo periodic aspect variations,
which leads to the observed flux variations.

The accretion flow powering coherent pulsations is variable in its own
right.  Like the wider class of LMXBs, AMXPs show aperiodic X-ray
variability over a wide range of timescales, including broad
band-limited noise structures and narrow quasi-periodic oscillations
(QPOs).  These aperiodic features have a rich structure of trends and
relations \citep{Straaten2005} that is nearly identical to the
variability observed in non-pulsating accreting neutron stars
\citep{Klis2006}.

Because the pulsations are powered by the accretion flow, it may be
expected that the periodic emission is coupled to the aperiodic
variability.  Considering the coherent and aperiodic variability
together gives a potential diagnostic of accretion physics.  Measuring
the degree of coupling, and exploring its dependence on energy or
luminosity, can give additional constraints on the physical models
aiming to explain quasi-periodic variability, the magnetic coupling of
the neutron star and the accretion flow, and the emission pattern of
the hotspot on the stellar surface.  Lacking a robust method for
analysing such coupling, however, previous studies have relied on
ad-hoc or model-dependent approaches.

Analysing observations of the canonical accreting millisecond pulsar
SAX J1808.4--3658, \citet{Menna2003} assume that the band-limited
noise is produced by a shot noise process and derive an analytic model
power spectrum for the scenario where the shot noise emission
originates partially from the neutron star hotspot.  This model could
be successfully fit to observations, which suggests that the periodic
and aperiodic components indeed show coupled behavior, but its results
are otherwise difficult to interpret.  This approach is complicated
further by the reliance on a shot noise model, which is ruled out by
the flux-dependent amplitudes that are observed in both the
periodic and aperiodic components of SAX J1808.4--3658
\citep{Uttley2001,Uttley2004}.

Considering the relation between an aperiodic flaring component and
the coherent pulsations in SAX J1808.4--3658, \citet{Bult2014}
selected the flares in the time domain and calculated the pulse
properties as a function of flux. While model independent, this
approach only works in rare cases where the aperiodic signal is
directly visible in the light curve. 

In a similar example, \citet{Ferrigno2017} measured pulse properties
over short time segments. Rather than using the flux distribution,
they computed the power spectrum for a series of consecutive pulse
amplitude measurements.  While this approach no longer requires the
aperiodic variability to be directly resolved in the time domain, it
does impose a minimum segment length in order to measure the pulse
properties through epoch folding.  Hence, only aperiodic features that
vary on timescales that are slower than the folding segment length can
be studied in this way.  Furthermore, this approach is difficult to
interpret as well, as the statistical properties of such a power
spectrum are not obvious. 

In this work we consider the problem of analysing the coupling between
the periodic and aperiodic variability of AMXPs, and propose a method
targeted specifically at quantifying such coupling. In section
\ref{sec:methods} we describe our analysis method in detail, and in
sections \ref{sec:results} and \ref{sec:discussion} we demonstrate its
use by applying it to observations of the accreting millisecond pulsar
IGR J00291+5943.

\subsection{IGR J00291+5934}
The transient X-ray binary IGR J00291+5934 (IGR J00291) is a 598.88 Hz
accreting millisecond pulsar in a 2.45 hour binary
\citep{Markwardt2004} located at a distance of $4.2\pm0.5$ kpc
\citep{DeFalco2017}.  This pulsar was discovered with
\textit{INTEGRAL} in December 2004 \citep{Eckert2004, Shaw2005}, when
it showed an outburst lasting approximately 15 days. In August and
September 2008 the source was again detected, showing two separate
outbursts each peaking at a flux level half that observed in 2004
\citep{Chakrabarty2008, Lewis2008}. The next major outburst, in July
2015, more closely resembled the behavior observed in 2004
\citep{DeFalco2017, Sanna2017b}.

The pulse profiles of IGR J00291 are nearly sinusoidal, with typical
sinusoidal fractional amplitudes of $8-12\%$ for the fundamental and
$\sim0.5\%$ or less for the second harmonic \citep{Galloway2005,
Falanga2005b}.  Tracking the arrival times of the pulsations showed
that the pulse frequency increases while the source is actively
accreting and decreases when it is not \citep{Patruno2010e,
Papitto2011}, which is interpreted as evidence for the accretion
torque spinning up the neutron star during outburst, and magnetic
dipole radiation causing a spin-down during quiescence. The pulse
phase was also observed to change linearly with X-ray flux
\citep{Patruno2010e}, likely indicating that the neutron star hotspot
moves over the stellar surface as the mass accretion rate changes
\citep{Patruno2009b}.

The aperiodic variability of IGR J00291 is somewhat anomalous as
compared to other AMXPs \citep{Linares2007, Hartman2011}. In both the
2004 and 2008 outbursts the power spectrum could be adequately
described with 3 to 4 broad noise components, together covering the
0.01--100 Hz range. The power spectral shape resembles that of a
typical atoll extreme island state (see, e.g.  \citealt{Doesburgh2017}
for a recent overview of the atoll states and their power spectra),
but has a much larger than usual total fractional variability of
$\sim50\%$ rms and lower than typical characteristic frequencies
ranging from about $0.04$ Hz up to $70$ Hz.  Additionally, two
harmonically related QPOs are observed at $\sim20$ mHz and $\sim40$
mHz that appear more like the low-frequency QPOs observed in black
holes \citep{Linares2007} than any feature normally seen in accreting
neutron stars.  The 2015 outburst showed a similar band-limited noise
structure, but different QPOs \citep{Ferrigno2017}.  In that outburst
a prominent $8$ mHz QPO was observed at low ($<2$ keV) energies.

Our choice for IGR J00291 as a test case is motivated by its high
pulse frequency and relatively large pulse fraction. The combination
of a high pulse frequency and lack of fast ($>100$ Hz) QPOs ensures
that there is little contaminating variability around the pulse
frequency in the Fourier domain, while a large pulse fraction
naturally contributes to a better significance in measuring the
coupling.

\section{Analysis Method}
\label{sec:methods}
    In this work we consider a simple premise: if the neutron star
    surface emission (i.e. the hotspot) couples to the variable
    accretion flow, then the pulse amplitude should, to some degree,
    exhibit the same aperiodic periodicities as that flow. 
    Our aim, then, is two-fold; first we want to define a Fourier
    transform of the pulsed emission to establish if the pulsations
    indeed show an aperiodically variable amplitude. If confirmed, the
    second step is to analyse if and how that aperiodicity correlates
    with the Fourier transform of the non-pulsed flux.

\subsection{Pulse modulation spectrum}
\label{sec:mod.spec}
    We define the pulse signal at time $t$ as
    \begin{equation}
    \label{eq:light.curve}
        x(t) = a_0(t) + a_1(t) \cos\rbr{2 \pi \nu_p t + \varphi_p},
    \end{equation}
    where $\nu_p$ is the pulse frequency, and $\varphi_p$ its phase
    offset. The function $a_0(t)$ gives the aperiodic non-pulsed
    emission, and $a_1(t)$ is the \textit{instantaneous amplitude},
    which is further specified as
    \begin{equation}
        a_1(t) = m_0 + m(t).
    \end{equation}
    Here $m_0$ gives the averaged pulse amplitude as measured by the
    usual coherent timing analyses \citep[see, e.g,][]{Hartman2008},
    and $m(t)$ is a generic zero-mean modulation function describing
    the stochastic variations about the mean amplitude. Our goal is to
    obtain the Fourier transform of $m(t)$.

    The last term in our expression for the pulse signal (eq.
    \ref{eq:light.curve}) is equivalent to an amplitude modulated (AM)
    carrier wave. The Fourier convolution theorem \citep{Bracewell1965}
    then implies that the spectrum of $m(t)$ appears in the Fourier
    transform of $x(t)$ as sidebands around the carrier frequency
    $\nu_p$.  This spectral information can be extracted if we
    \textit{demodulate} the light curve, which amounts to applying a
    frequency shift in the Fourier domain. Applying the Fourier shift
    theorem \citep{Bracewell1965}, we write the demodulated light
    curve as
    \begin{equation}
    \label{eq:demodulation}
        z(t) = x(t) e^{-2 \pi i \nu_p t - i \varphi_p},
    \end{equation}
    so that the Fourier transform at frequency $f$ is given as
    \begin{align}
        Z(f)
        &= \int_{-\infty}^{\infty} z(t) e^{-2 \pi i f t} dt, \nonumber\\
        &= A_0(f+\nu_p) e^{-i\varphi_p} 
         + \frac{1}{2} \Big[ m_0 \delta(f) + M(f) \Big] \nonumber\\
        &+ \frac{1}{2} \Big[ m_0 \delta(f+2\nu_p) + M(f+2\nu_p) \Big]
         e^{-i 2\varphi_p},
    \end{align}
    where $\delta(f)$ gives the Dirac delta function and upper case
    letters are used to indicate the Fourier transforms of lower case
    time domain functions. We see that modulation spectrum, $M(f)$,
    makes two contributions; once centered at zero frequency and once
    centered at $-2\nu_p$.  These two
    contributions have an Hermitian symmetry\footnote{
        Hermitian symmetric functions have the symmetry relation $H(f)
        = H^\ast(-f)$, where $\ast$ denotes the complex conjugate. In
        our case the symmetry is about $f=-\nu_p$ rather than $f=0$
        due to the applied frequeny shift operation. 
    }, about frequency $-\nu_p$
    so we can capture all information with only one of these two
    contributions.  The spectrum of the non-pulsed emission, $A_0(f)$, acts
    as a contaminating background term. If this spectrum is
    band-limited to cut off well below the pulse frequency, then we
    can effectively separate all terms in frequency space, and write
    \begin{equation}
    \label{eq:demod.transform}
        Z(f) = \frac{1}{2} M(f),
    \end{equation}
    for $0 < \abs{f} \ll \nu_p$.
    This is a fair assumption for IGR J00291, but does not
    hold in general. We therefore caution that for AMXPs that show
    aperiodic variability at frequencies comparable to $\nu_p$
    there may be some contamination in eq. \ref{eq:demod.transform}
    from the high-frequency tails of $A_0(f)$  that should be
    accounted for.

    The pulse modulation spectrum, $M(f)$, is not necessarily
    symmetrical around zero frequency. Depending on how, exactly, the
    periodic emission is modulated, this spectrum may show features at
    negative frequencies, at positive frequencies, or both. 
        For instance, the beat frequency between two azimuthal
        rotations would cause a strictly one-sided modulation
        spectrum, whereas a surface temperature oscillation could
        cause a symmetrical modulation spectrum.
    To proceed, we therefore extract separate modulation sidebands for
    both the positive or negative frequencies.  

\begin{figure}[t]
        \centering
        \includegraphics[width=1.0\linewidth]{{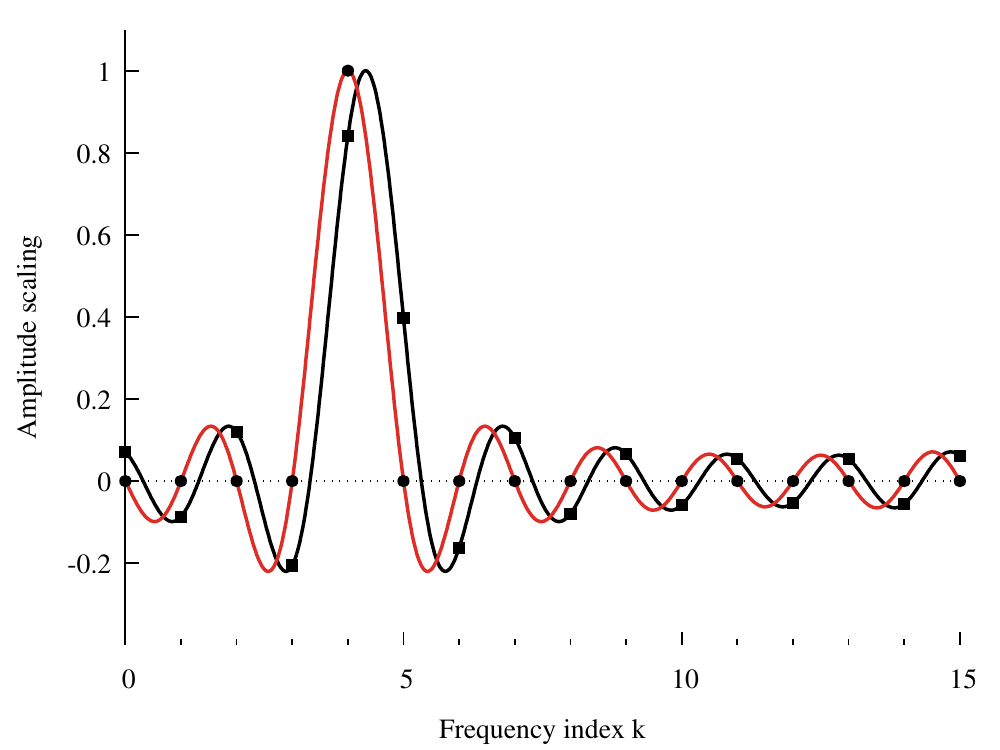}}
        \caption{%
            Discrete-Time Fourier Transform of a complex wave
            for $N=16$ and $\kappa=4$ (red, circles) and $\kappa=4.32$
            (black, squares). The curves show the Dirichlet kernel response
            and the symbols mark the samples of the usual Discrete Fourier 
            Transform (DFT). Note that for $\kappa=4$ the DFT samples
            coincide with the kernel zero-crossings, thus minimizing
            spectral leakage.
        }
        \label{fig:dirichlet}
    \end{figure}

    A known difficulty in the analysis of pulse sidebands is that
    observations are discretely sampled, so that  
    windowing and spectral leakage cause spurious broadening of the
    pulse spike \citep{Lazzati1997,Burderi1997}. Both phenomena can
    be described at once by considering that the Discrete-Time Fourier 
    Transform (DTFT) can be written as a convolution \citep{Bloomfield1976}
    \begin{align}
        \widetilde X(f)
        &= \sum_{j=0}^{N-1} x(j\Delta t) e^{- 2 \pi i j \Delta t f}, \nonumber \\
        &= X(f) \otimes D_N(f),  \label{eq:dtft}
    \end{align}
    where we use $\otimes$ to indicate the convolution. The
    light curve is measured at $N$ discrete intervals with time
    resolution $\Delta t$ and 
    \begin{equation}
        D_N(f) = 
            \frac{\sin(\pi f / \Delta f)}{\sin(\pi f / N \Delta f)} 
            e^{i(N-1) \pi f / N \Delta f }
    \end{equation} 
    is a version of the Dirichlet kernel \citep{Titchmarsh1939}. We
    also used $\Delta f = 1/N\Delta t$ so that the more familiar
    Discrete Fourier Transform (DFT) is obtained by evaluating eq.
    \ref{eq:dtft} only at the discrete Fourier frequencies $k \Delta
    f$, where $k=-N/2, \dots, 0, \dots, N/2-1$.
    For a complex waveform $x_j = e^{2 \pi i j \kappa / N}$ with
    frequency $f_{\rm wave} = \kappa\Delta f$, we get 
    the DTFT
    \begin{equation}
        \widetilde X_{\rm wave}(f) = D_N((k-\kappa) \Delta f),
    \end{equation}
    which we show in Figure \ref{fig:dirichlet} for two values
    of $\kappa$.     
    It can be seen that the wave frequency parameter $\kappa$ acts to
    translate the response of the Dirichlet kernel. The observed
    spectral leakage in the DFT then arises if $\kappa$ is not an
    integer.
    Notice, however, that the demodulation of the light curve (eq.
    \ref{eq:demodulation}) acts to precisely align the Fourier transform
    with the pulse frequency. The Fourier frequency samples then
    coincide with the zero crossings of the Dirichlet kernel, ensuring
    that the pulse spike will not be broadened outside the zero
    frequency bin of the DFT.

    An additional distortion comes from the fact that the discrete
    measurements reflect an integration over the width of a time bin
    rather than an instantaneous sample as assumed above. This effect
    can be described by convolving the time domain function with a
    binning window \citep{Klis1989}. In the frequency domain this
    distortion causes slightly suppressed amplitudes, but it will not
    bias the phases. Additionally, if the amplitude of the coherent
    wave is modulated, then the modulation spectrum itself will still
    suffer from sampling effects, however, those effects will be
    identical to the ones affecting the low frequency range of the
    regular power density spectrum.

    Writing the discrete version of the modulation spectrum 
    \begin{equation}
    \label{eq:mod.spec}
         M_k = 
        \left\{
            \begin{array}{l l}
                2 Z_k,      & \mbox{\it positive frequency sideband,} \\
                2 Z_{-k}^*, & \mbox{\it negative frequency sideband,} \\
            \end{array}
        \right.
    \end{equation}
    we obtain the Leahy-normalized modulation power spectrum \citep{Leahy1983a} as
    \begin{equation}
    \label{eq:mod.leahy.pds}
        P_{k, {\rm Leahy}} = \frac{1}{2 N_\gamma} \abs{  M_k }^2,
    \end{equation}
    where $N_\gamma$ gives the number of photons in the time series.
    For counting noise data the resulting powers are $\chi^2$
    distributed for $2$ degrees of freedom, and can be treated on the
    same footing as the regular power density spectrum \citep[see,
    e.g,][]{Klis1989}.

    While the Leahy(-like) normalization is useful to determine the
    uncertainties on the computed powers, it is more informative to consider
    the rms normalization. We express the powers in units of rms as a
    fraction of the mean pulse amplitude, rather the mean intensity. The
    squared fractional rms normalization can be obtained by scaling the
    Leahy(-like) powers as
    \begin{equation}
    \label{eq:mod.rms.pds}
        P_{k, {\rm rms}} = \frac{4}{a^2} \frac{T}{N_\gamma} P_{k},
    \end{equation}
    where $T=N\Delta t$ gives the duration of the time series, $a =
    m_0 / \avg{a_0}$ is the fractional sinusoidal amplitude of the
    pulsation, with $\avg{a_0}$ the time average of $a_0(t)$.

    If we find that the two sidebands produce statistically consistent power
    spectra, and that they form a complex conjugate pair in the
    complex plane, then we have demonstrated the modulation spectrum
    is Hermitian about zero frequency. This immediately puts a 
    constraint on which physical processes might cause the modulation.
    Additionally, we can exploit this symmetry by treating each of the
    two sidebands as independent samples of the same underlying
    spectrum. This allows us to improve the statistical quality of
    the modulation spectrum by averaging the sidebands in the complex
    plane before computing the modulation power density spectrum.

\subsection{The modulation coherence}
\label{sec:mod.coherence}
    If the stochastic variation of the pulse amplitude, $m(t)$, is
    coupled to aperiodic variability, then the non-pulsed emission
    term $a_0(t)$ should also have some dependence on that modulation. In
    the Fourier domain this results in a phase relation between $M(f)$
    and the low frequency part of the normal spectrum, $A(f)$, that
    can be studied using the cross-spectrum \citep{Jenkins1968}.

    Exactly how the modulation term enters into the non-pulsed
    emission depends on the assumed accretion geometry and the various
    origins of the emission components. For instance, one may assume
    that a part of the hotspot is alway in the observer's
    line-of-sight, so that $m(t)$ contributes directly to $a_0(t)$.
    Through emission from the accretion disk, we may also be able to
    observe the modulation term directly, possibly at a time-lag with
    respect to the pulse modulation. Additionally, either of these two
    contributions may be non-linear. 
    
    In this work, however, we take a first order approach and consider
    the direct Fourier transform of the light curve, $X(f)$, as an
    estimator of $A(f)$ and proceed to analyse its linear correlation
    with $M(f)$. Hence, we define the complex-valued cross spectrum 
    \begin{equation}
        C(f) = X(f)  M^*(f),
    \end{equation}
    where $ M(f)$ is given by eq. \ref{eq:mod.spec}.
    Casting the cross spectrum in complex polar form, we can extract the
    co-amplitude spectrum by taking the absolute value of $C(f)$. To measure
    the degree of coupling it is more convenient to normalize the co-amplitudes
    to a coherence measure \citep{Vaughan1997}
    \begin{equation}
    \label{eq:mod.coherence}
        \abs{\gamma}^{2} 
            = \frac{\abs{\avg{C}}^{2}-n^{2}}{\avg{P_{X}}\avg{P_{M}}},
	\end{equation}
	where 
    \begin{align}
        n^2 
        &= \frac{1}{K} 
        \Big\{ \avg{P_X} P_{{\rm noise},M} + \avg{P_M} P_{{\rm noise},X} \nonumber \\ 
        &- P_{{\rm noise},X} P_{{\rm noise},M} 
           \Big\}
    \end{align}
    corrects for the bias due to the presence of Poisson counting noise power, 
    $P_{\rm noise}$, and $\avg{\cdot}$ represents an
    ensemble average over $K$ independent light curve segments and/or
    adjacent frequency bins.
    Finally we can consider the phase of the cross spectrum, $\phi(f)$, 
    to establish whether the pulse modulation is leading or lagging behind 
    $X(f)$.

\section{Data Reduction}
\label{sec:datareduction}
    We analyse observations of all observed outbursts of IGR J00291. For the
    2015 outburst we consider an \xmm observation. For the outbursts
    of 2008 and 2004 we consider observations with the \textit{Rossi X-ray
    Timing Explorer} (\rxte).

\subsection{XMM Newton}
    The 2015 outburst of IGR J00291 was observed with \xmm on July 28 (ObsID
    0744840201). We analyze the \textsc{epic-pn} data, which was collected in
    \textsc{timing} mode, providing event data at a time resolution of $29.56$
    $\mu$s.

    We process the \xmm data with \textsc{sas} version 15.0.0, using the most 
    recent calibration files available. Standard screening criteria were
    applied, selecting only those events in the energy range $0.4-10$ keV with
    $\textsc{pattern}\leq4$ and screening $\textsc{flag}=0$. The 
    source events were extracted from the \textsc{rawx} coordinates $\br{34:43}$. 
    Likewise, the background count rate was extracted from the $\br{3:5}$ range. 

    The extracted event arrival times were corrected to the Solar System barycenter
    using the \textsc{barycen} tool, based on the source coordinates of the
    optical counterpart \citep{Torres2008}.
    Subsequently the arrival times were adjusted for the binary motion of the
    neutron star based on the ephemeris of \citet{Sanna2017b}.

\subsection{RXTE}
    We analyze all pointed \rxte observations of the 2008 (Proposal number
    93013 / 93435) and the 2004 (Proposal number 90052 / 90425) outbursts of
    IGR J00291.  We consider all GoodXenon and (122 $\mu$s) Event data,
    applying standard screening criteria (source elevation above $10^\circ$ and
    pointing offset less than $0.02^\circ$) and selecting only those events in
    energy channels 5 to 48 ($\sim 2-20$ keV).
	
    Using the source coordinates of \citet{Torres2008} we correct the photon
    arrival times to the Solar System barycenter using the \textsc{ftool}
    \textsc{faxbary}, which also applied the \rxte fine clock corrections. We
    then use the binary ephemeris reported by \citet{Patruno2010e} to correct
    the data for the binary motion of the neutron star. 

    The \rxte data is contaminated by the foreground emission of the nearby
    Intermediate Polar V709 Cas \citep{Falanga2005b}. Because this source does
    not contribute to the variability above $5$ mHz \citep{Linares2007} we can
    treat it as an additional contribution to the background emission. We 
    use the \textsc{ftool} \textsc{pcabackest} to estimate the instrumental
    background level and correct for estimated count rate of V709 Cas
    \citep{Linares2007,Hartman2011}.

\subsection{Timing}
    For both the \rxte and \xmm observations we bin the data to a time
    resolution of $\sim240$ $\mu$s and divide the light curves into segments of
    $\sim1000$ s. Fourier transforms of these light curve segments then have a
    frequency resolution of $\sim 1$ mHz and a limiting Nyquist frequency of
    $\sim2000$ Hz.
    
    To construct the pulse amplitude modulation power spectra we first
    fold each data segment on the pulse period and measure the local
    pulse phase and amplitude.  We then demodulate the time series
    using eq. \ref{eq:demodulation} and use eq. \ref{eq:mod.leahy.pds}
    to construct the power spectra.  For the \xmm data we average all
    spectra in the ObsID, while for the \rxte observations we use the
    data grouping of \citet{Linares2007} and \citet{Hartman2011}. 
    
    Because the shifting operation of eq. \ref{eq:demodulation} will
    affect the dead time process in a non-trivial way, the Poisson noise
    model of \citet{Zhang1995} may not be appropriate. However, given
    that the considered count-rates are relatively low, dead time
    effects should be weak so that the Poisson noise power is well
    approximated by a constant.  Following the approach of 
    \citet{KleinWolt2004}, we therefore fit a constant to the
    average power at high frequencies, which we then adopt as the
    Poisson noise level. 
   
    The Poisson noise-corrected spectra are renormalized to fractional
    rms with respect to the pulse amplitude (eq. \ref{eq:mod.rms.pds})
    and fitted with a multi-Lorentzian model
    \citep{Belloni2002}, where each Lorentzian, $L\rbr{ \nu ; r, Q, \nu_{\rm
    max}}$, is expressed in terms of its integrated rms amplitude
    $r$, where
    \begin{equation}
        r^2 = P = \int_0^{\infty} L(\nu) d\nu,
    \end{equation}
    its characteristic frequency $\nu_{\rm max}$, and quality factor $Q$. A
    component is accepted as significant if its integrated power, $P$, has a
    single-trial signal to noise ratio greater than three; that is, when
    $P/\sigma_P \geq 3$.

\section{Results} 
\label{sec:results}

    In the following we apply the methods discussed in section
    \ref{sec:methods} to the observations of IGR J00291. We first present the
    results from the \xmm data.  This data contains a strong $8$ mHz QPO that
    was previously shown to couple to the pulsations in a heuristic analysis
    \citep{Ferrigno2017}, and hence provides an opportunity validate our
    method.
    Next we present the results from the \rxte data, for which low frequency
    QPOs have been reported \citep{Linares2007,Hartman2011}. These weak
    features may not be resolved directly in the light curve, and require a
    method as developed here to establish whether they too couple to the pulsed
    emission.

\subsection{XMM-Newton}

\begin{figure}[t]
    \centering
    \includegraphics[width=\linewidth]{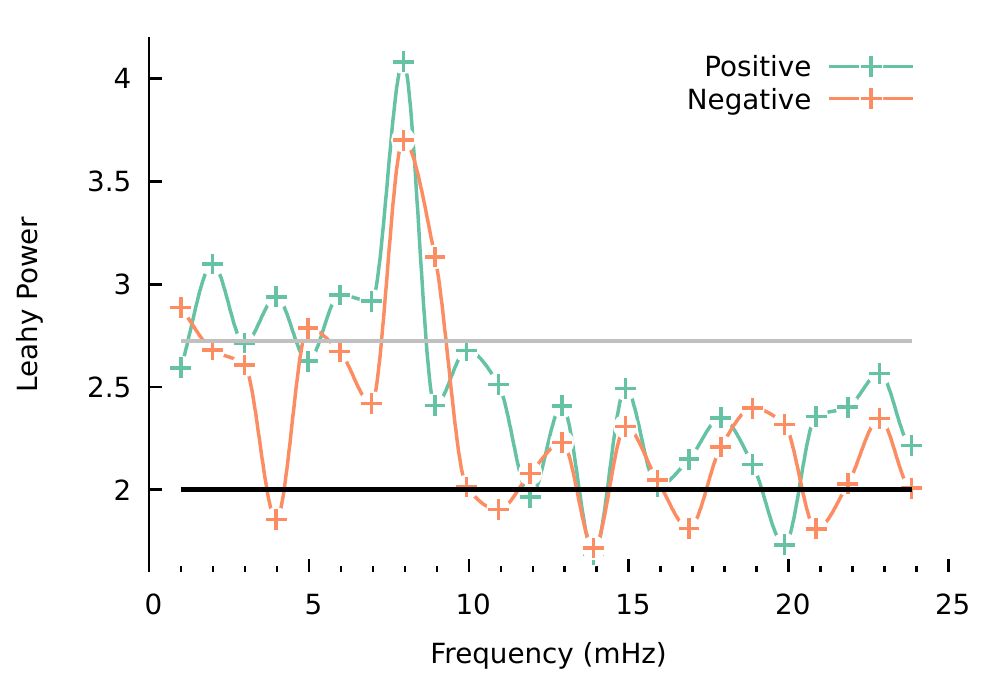}
    \caption{%
        Pulse modulation power spectra of the \xmm observation,
        showing both the positive and negative frequency sidebands.
        The black line indicates the Poisson noise power level
        and the grey line show the $3\sigma$ detection threshold.
    }
    \label{fig:xmm.sidebands}
\end{figure}

\begin{figure}[t]
    \centering
    \includegraphics[width=\linewidth]{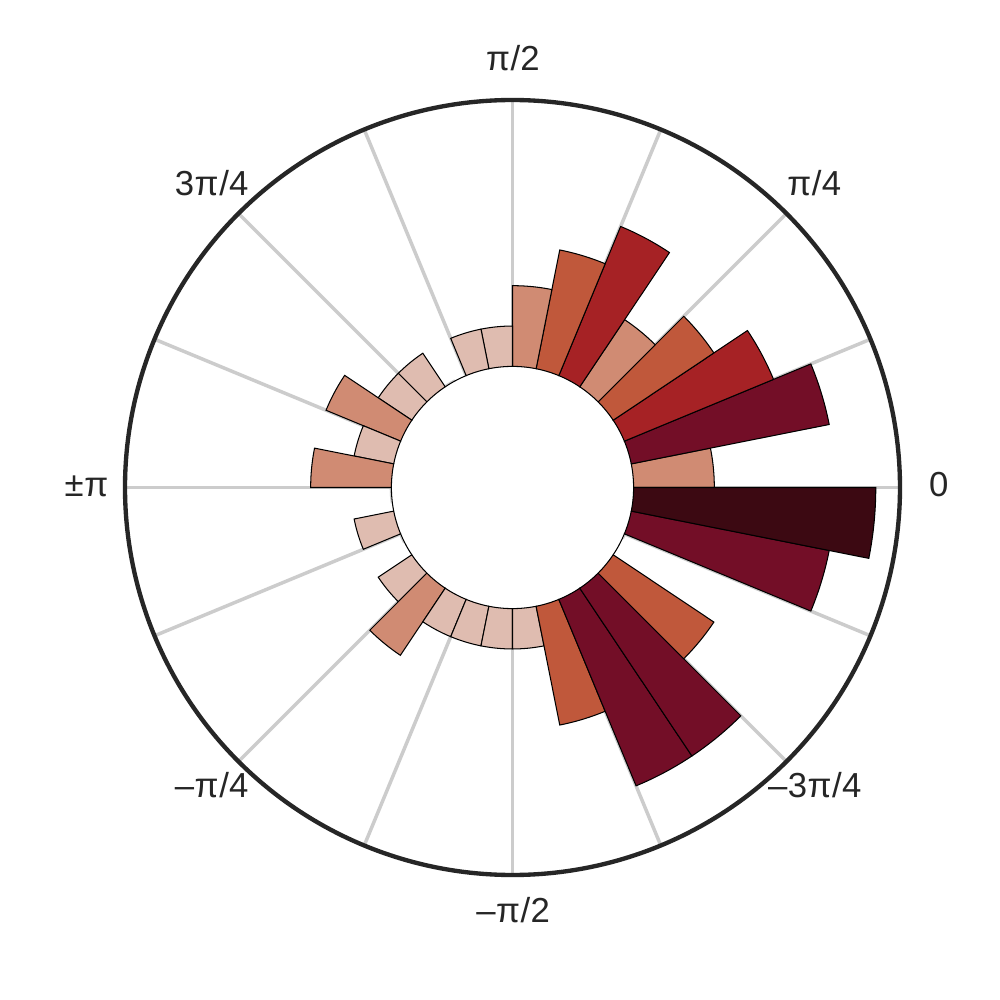}
    \caption{%
        Histogram of the phase difference between the positive sideband
        modulation and the complex conjugate of the negative sideband modulation
        in the \xmm observation, when both sidebands are evaluated at $8$ mHz.
        The colors scale with the probability density such that darker
        colors phase bins indicate a larger number of occurrences in
        that direction. The average phase is consistent with zero
        ($\phi=0.1\pm1.4$),
        indicating the two sidebands are symmetric. 
    }
    \label{fig:xmm.phase.distribution}
\end{figure}

    To confirm that the 8 mHz QPO is coupled to the pulsed emission, we
    first calculate the modulation power spectrum for the positive and
    negative frequency sidebands of the pulsations.  The lowest
    frequency samples of these spectra are shown in Figure
    \ref{fig:xmm.sidebands}.  Both sidebands have a significant feature
    at $8$ mHz.  The sidebands are statistically consistent with each
    other, suggesting they are the symmetric wings of a single
    modulating function.  If so, then the negative sideband should be
    the complex conjugate of the positive sideband (section
    \ref{sec:mod.spec}).  We can test this relation by measuring the
    complex phase angle at $8$ mHz in each sideband, and then exploring
    the distribution of their sum. If the two features form a conjugate
    pair, then the summed phase should have, on average, zero phase.
    This phase distribution (shown in Figure \ref{fig:xmm.phase.distribution})
    indeed has a preferential direction that is statistically consistent
    with being zero to within 20\% uncertainty ($\phi = 0.1 \pm 1.4$),
    confirming that there is a single modulation mechanism that is
    responsible for both sidebands.

\begin{figure}[t]
    \centering
    \includegraphics[width=\linewidth]{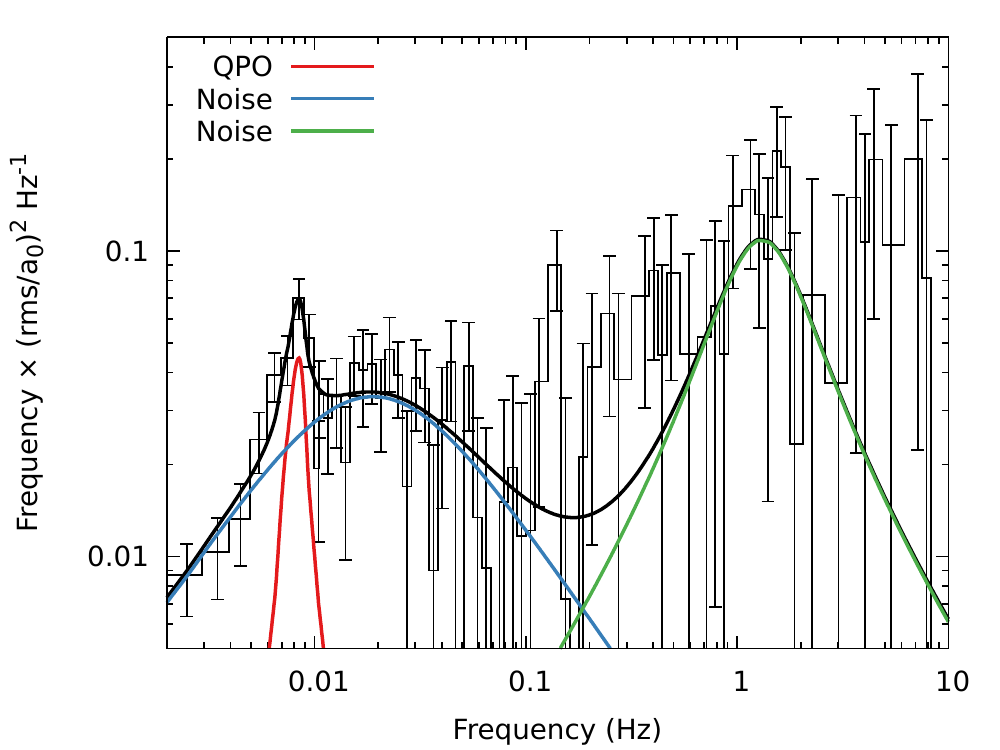}
    \caption{%
        Pulse modulation power spectra in the $0.4-2$ keV energy band of the 
        \xmm observation computed by averaging the positive and
        negative sideband. The curves give the best-fit model
        reported in Table \ref{tab:fit.results}.
    }
    \label{fig:pulse.mod.spectrum}
\end{figure}

    We proceed to construct the averaged-sideband pulse amplitude
    modulation power spectrum. That is, for each segment we extract
    the modulation spectrum by averaging the two sidebands in the
    complex plane.  This averaged-sideband spectrum (Figure
    \ref{fig:pulse.mod.spectrum}) shows a narrow QPO at $8$ mHz with a
    fractional amplitude of $12\%$ rms with respect to the mean pulse
    amplitude. The QPO is superimposed on a broad band-limited noise
    term with a characteristic frequency of $0.020$ Hz, and a second
    noise term is found centered at $1$ Hz.  The detailed fit
    parameters are given in Table \ref{tab:fit.results}.

\begin{table*}
    \centering
    \newcommand{\mc}[1]{\multicolumn2c{#1}}
    \newcommand{\mr}[2]{\multirow{#1}{*}{#2}}
    \newcommand{\customhead}[1]{\tableline \multicolumn{9}{c}{#1} \\ \tableline}
    \caption{%
        Pulse amplitude modulation power spectrum fit results
        \label{tab:fit.results}	
    }
    \begin{tabular}{ l l D D D c }
        \decimals
        \tableline
        Group & Component & \mc{Frequency} & \mc{Quality} & \mc{Amplitude} & $\chi^2$ / dof \\
              &           & \mc{(Hz)}      & \mc{~}       & \mc{(\% rms)}  &          \\
        \customhead{2015}
        ~   & QPO       & 0.0082(2) &  5. (2)      &  12.1(1.8) &  ~ \\
        XMM & noise$_1$ & 0.019(3)  &  0. ~(fixed) &  32.4(1.5) &  74 / 75 \\
        ~   & noise$_2$ & 1.3(3)    &  0.7(4)      &  41. (6)   &  ~ \\
        \customhead{2004}
        ~   & break   &    0.36(13)  &  0. ~(fixed) &  28.7(3.0) & ~       \\
        A1  & QPO$_1$ &  0.0195(14)  &  1.7(6)      &  11.7(1.4) & 60 / 63 \\
        ~   & QPO$_2$ &  0.0445(16)  &  3.7(1.9)    &  10.8(1.8) & ~       \\
        \tableline
        \mr{2}{A2} & break & 0.9(5)   &  0. ~(fixed)  &  46.1(7.6) & \mr{2}{53/59} \\
                   & noise & 0.030(7) &  0.20(18)     &  32.3(3.1) & \\
        \tableline
        B   & noise & 0.031(9)  &  0.2(3)  &  60.7(6.7) & 44/52 \\
        \customhead{ 2008 }
        P1  & noise & 0.024(4)   &  0. ~(fixed)  &  92.3(58) & 37 / 41 \\
        \tableline
        P2  & noise & 0.067(18)  &  0. ~(fixed)  &  95.8(90) & 60 / 62 \\
        \tableline
    \end{tabular}
\end{table*}

\begin{figure}[t]
    \centering
    \includegraphics[width=\linewidth]{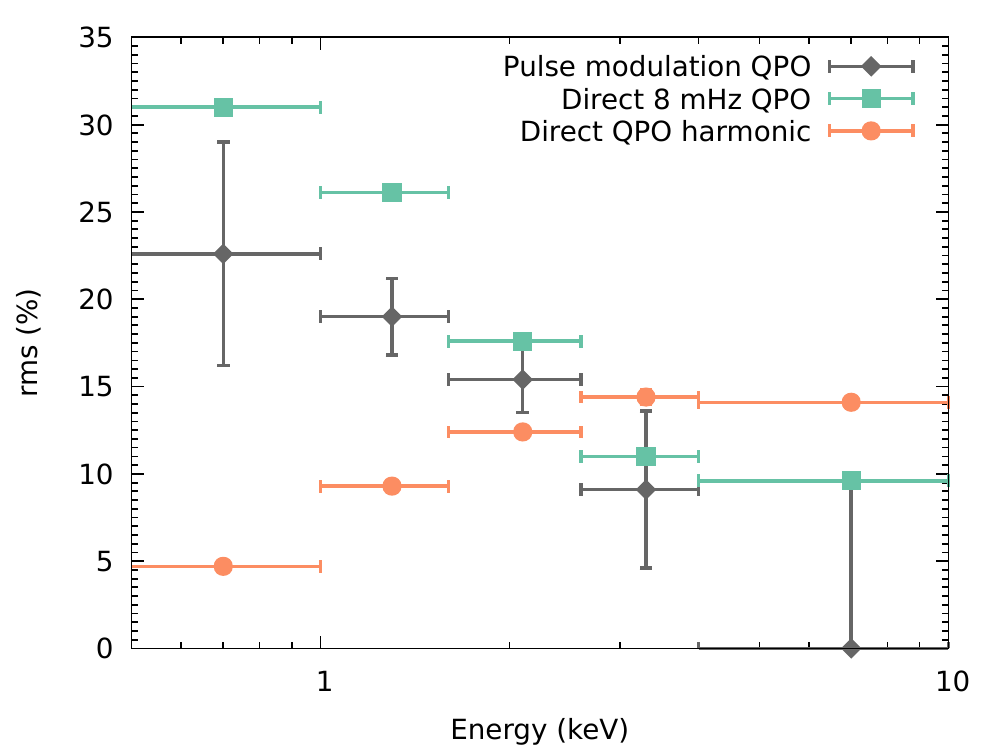}
    \caption{%
        Fractional amplitude of the $8$ mHz QPO as a function of energy,
        showing the QPO as measured in the regular power spectrum (green squares),
        and as measured in the modulation power spectrum (grey diamonds). Also
        shown is the amplitude of the QPO harmonic (orange circles) as detected 
        in the regular power spectrum. The pulse modulation QPO is
        measured in terms of fractional rms with respect to the pulse
        amplitude, whereas the direct QPO and its harmonic are
        expressed in terms of fractional rms with respect to the mean
        flux.
    }
    \label{fig:rms.energy}
\end{figure}

    To explore the energy dependence of the QPO, we construct the
    modulation power spectrum for different energy bands. As shown in
    Figure \ref{fig:rms.energy} the fractional amplitude of the QPO
    peaks at $1$ keV and drops rapidly with increasing energy. Above
    $4$ keV the QPO could no longer be resolved in the modulation
    power spectrum.  For comparison we have also measured the
    fractional amplitude of the direct $8$ mHz QPO and its harmonic
    component as observed in the regular power spectrum using the same
    energy bands. We find that both the direct $8$ mHz QPO and the
    pulse amplitude modulation QPO show roughly the same energy
    dependence, and that the pulse modulation has a systematically
    lower fractional amplitude.
    Remarkably, the harmonic of the $8$ mHz QPO as detected in the
    direct power spectrum has an opposite trend, showing an increasing
    amplitude for higher energy. Additionally, this harmonic is only
    detected in the direct power spectrum, and not in modulation power
    spectrum.

\begin{figure}[t]
    \centering
    \includegraphics[width=\linewidth]{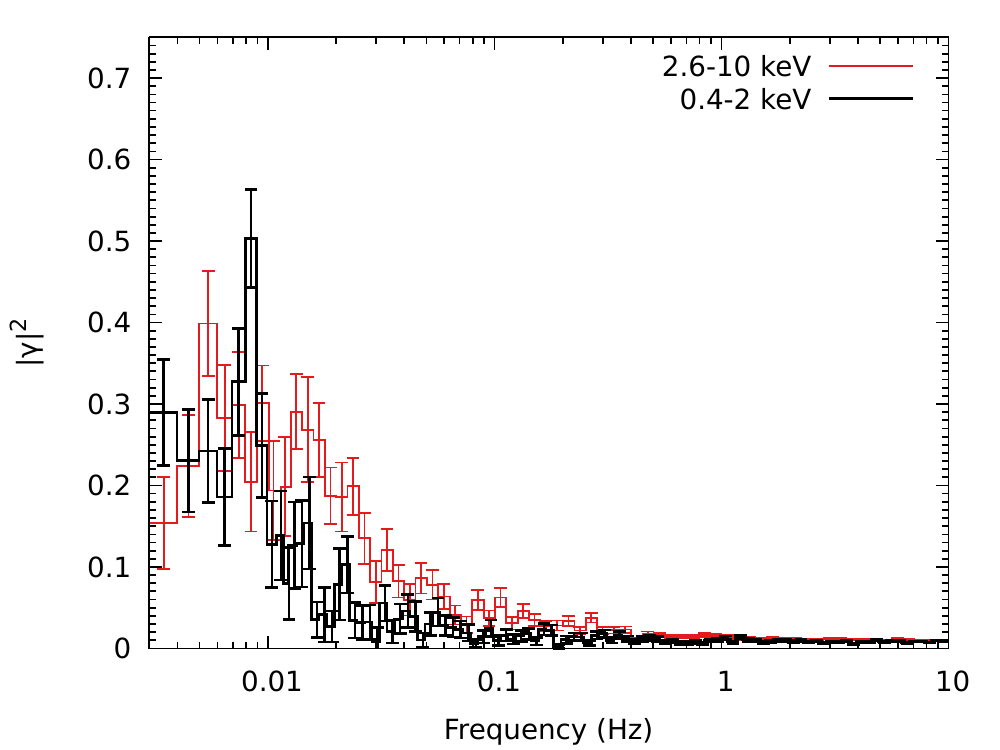}
    \caption{%
        Pulse amplitude coherence spectrum in the $0.4-2$ keV (black)
        and $2.6-10$ keV (red) bands of the \xmm observation.
    }
    \label{fig:xmm.coherence.spectrum}
\end{figure}

    In order to conclusively prove that the $8$ mHz QPO indeed couples
    to the pulsed emission, we compute the modulation coherence spectrum
    (eq \ref{eq:mod.coherence}) for the \xmm observation, which we show
    in Figure \ref{fig:xmm.coherence.spectrum}. The spectrum shows a
    clear peak at $8$ mHz, where the magnitude-squared coherence measure
    is $\sim 0.5$.  
    Considering the phase distribution of the cross spectrum, we find
    that the associated phase-lag at 8 mHz is consistent with being
    zero to within a $10\%$ uncertainty, indicating the 8 mHz QPO in
    the pulse modulation is in phase with the corresponding feature in
    the direct power spectrum.
    The coherence spectrum
    shows additional peaks at $16$ mHz and $24$ mHz, each marginally
    detected with $3\sigma$ significances and having
    progressively smaller coherence measures. For these higher
    harmonics, however, the cross spectral amplitudes are too low to
    reliably measure the phase lag. If we construct the modulation
    coherence spectrum for higher energies, we find that, consistent
    with the direct power spectrum, the $8$ mHz features becomes less
    pronounced. Meanwhile the coherence measure at the harmonic frequencies
    increases, even though the peaks themselves can longer be
    individually resolved.

\subsection{RXTE}

    Following the nomenclature of \citet{Linares2007}, the \rxte observations
    of the 2004 outburst are divided into three groups; set A1, set A2, and set B. 
    For each set we calculate the modulation power spectrum, using a bandwidth 
    of 200 Hz. The spectra show no features above 10 Hz, so we use the 20 Hz to
    200 Hz range to compute the Poisson noise level and subsequently truncate
    the spectra above 10 Hz before proceeding with the multi-Lorentzian fit.

    Following the same procedure as in the previous section, we first construct
    the pulse amplitude modulation power spectra for the positive and negative
    sidebands of the pulse spike separately. We find that both spectra show a
    broad noise structure at mHz frequencies, with individual peaks on top of
    them. With a significance of $\sim2\sigma$, these peaks are not formally
    detections, however, they have coincident frequencies between the two
    sideband spectra. 
    
\begin{figure}[t]
	\centering
    \includegraphics[width=1.0\linewidth]{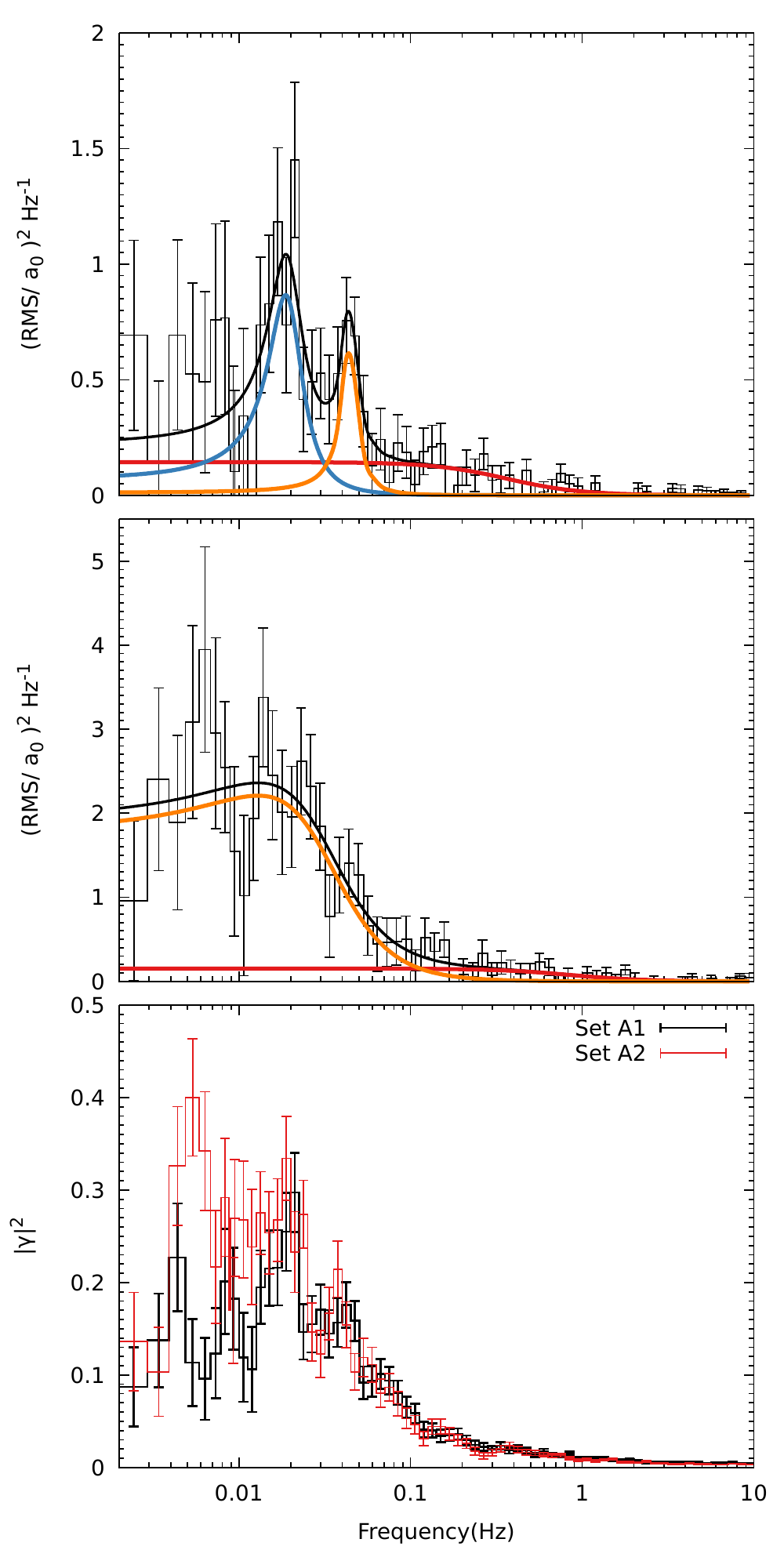}
	\caption{%
        Pulse amplitude modulation power spectra for set A1 (top panel) and A2
        (middle panel) of the 2004 outburst with their best fit
        multi-Lorentzian models, and the modulation coherence spectra of
        both sets (bottom panel). See Table \ref{tab:fit.results} for the model
        fit parameters.
	}
	\label{fig:mspec.a1}
\end{figure}

    If the two sidebands represent the same underlying spectrum, then using the
    averaged sideband should boost the signal-to-noise of these features. We
    therefore proceed by constructing the modulation power spectra for the
    averaged sideband, in which we indeed detect statistically significant QPOs.
    The spectrum is shown in the top panel of Figure \ref{fig:mspec.a1}, and
    can be described with two QPOs at characteristic frequencies of $20$ mHz
    and $44$ mHz, and an additional broad noise component with a characteristic
    frequency of $0.4$ Hz (see Table \ref{tab:fit.results} for details). 
    These QPOs are consistent with those reported by \citet{Linares2007} for
    the direct power spectrum, although the fractional amplitude of the
    modulation QPOs is twice that of the direct QPOs.
	
    For set A2 we again find that a broad noise component is present in
    the positive and negative sidebands of the pulse spike, and construct
    the pulse amplitude modulation spectrum for the averaged sideband.
    The resulting spectrum again shows features at $20$ and $44$ mHz, however,
    they could not be individually resolved. Instead we describe the 
    power in that frequency range using a single broad noise component (see
    Figure \ref{fig:mspec.a1} middle panel). The spectrum also shows an
    additional excess of power at a frequency of $6$ mHz, however, with 
    a $2\sigma$ detection level this feature is not statistically
    significant.

    Finally, set B covers the tail of the outburst, where both the source count
    rate and the pulse fraction have decreased (see \citet{Mukherjee2015} for
    the pulse fraction as a function of time). Because the signal-to-noise
    is much lower, we had to smooth the spectra over wide frequency bins to
    be able to detect the power. A single broad noise component provided
    an adequate fit to the averaged sideband modulation power spectrum.
    This noise term has a similar shape and frequency as the mHz noise term
    seen in set A2, but shows a higher fractional rms.

    Following section \ref{sec:mod.coherence}, we also calculated the modulation
    coherence spectra for the three sets. The coherence spectra of all sets
    show a broad structure below $0.1$ Hz, and for sets A1 and A2 a number of
    peaks are observed (see Figure \ref{fig:mspec.a1}). For set B the required
    smoothing prohibited the detection of such narrow features. 
    In the case of set A1 there are distinct peaks in the coherence measure at 
    roughly $20$, $40$ and $60$ mHz, indicating that mHz QPOs are coupled to the 
    pulsations. The modulation coherence spectrum of set A2 shows similar peaks, 
    suggesting the presence of a coupled component despite the modulation power 
    spectrum being poorly resolved. 
    Additionally, a prominent peak exists in the modulation coherence spectrum 
    at $6$ mHz, which would suggest that the power excess at $6$ mHz that was
    seen in the modulation power spectrum is a real component.

    Following the analysis of \citet{Hartman2011}, we group the \rxte
    observations of the 2008 outbursts in two sets; one for each partial
    outburst, and perform the same analysis. The modulation power spectra of
    these two outbursts are very similar (see Table \ref{tab:fit.results} for
    the fit parameters). Although a number of very narrow features may be seen
    in the spectra, none are statistically significant. Instead both spectra
    are adequately described with just a single broad noise component. Those
    noise terms, however, do have a remarkably large amplitudes, showing
    fractional amplitudes in excess of $90\%$ rms with respect to the pulse
    amplitude. The modulation coherence spectra show a broad structure as well,
    with a magnitude squared coherence measure of $\sim0.25$ below $30$ mHz,
    and no resolved peaks.

\section{Discussion}
\label{sec:discussion}

Using specialized methods introduced in this work, we have analysed the coupling
of the millihertz QPOs in IGR J00291 with its 599 Hz pulsations.  We can
distinguish two instances of mHz QPOs, namely the $8$ mHz QPO observed with
\xmm in the 2015 outburst \citep{Ferrigno2017} and the $\sim20$ mHz and
its harmonics as seen in \rxte observations of earlier outbursts \citep{Linares2007,
Hartman2011}. Because the two versions of millihertz variability are not
observed simultaneously, nor in the same energy band, it is unclear if they
share the same underlying physical mechanism \citep{Ferrigno2017}.
In the following we therefore first discuss the \xmm and \rxte results
separately, before briefly considering them together. 

\subsection{The 8 mHz QPO}
By analysing the sidebands of the pulsations we have shown that
the pulsed emission couples to the $8$ mHz QPO. The fractional
modulation of the pulsed emission was found be systematically lower
than that of the direct QPO.
Further confirmation that the pulsed emission and the QPO are coupled
comes from the detection of a $0.5$ coherence measure between them.

Interestingly, the fractional amplitude of the pulse modulation
shows the same energy dependence as the direct QPO, with a peak
amplitude at $1$ keV, and decreasing amplitudes for increasing
energy. By contrast, the average pulse amplitude has an opposite
energy dependence; the pulse fraction is low at $1$ keV and
increases steeply over the same energy range \citep{Sanna2017b}.
Meanwhile, the harmonic of the direct QPO also shows an increasing
fractional amplitude with energy. Although the modulation power
spectrum does not show a clear harmonic, this trend of an increasing
harmonic and decreasing fundamental as a function of energy is also
observed in the coherence spectrum. 

The pulsations of AMXPs are typically well described by a two component
spectral model consisting of a soft thermal blackbody and a hard power-law
tail \citep{Gierlinski2002, Poutanen2003}. Both components are associated with
the impact of the accretion flow, which is thought to cause a hotspot on the
stellar surface and an accretion shock slightly above it. The blackbody is then
attributed to the hotspot itself, and the power-law to the accretion shock, which
reprocesses the hotspot emission by thermal Comptonization
\citep{Gierlinski2005}.

The natural explanation for the pulse/QPO coupling would be that the
QPO emission originates in the stellar hotspot. This assumption,
however, does impose some restrictions on how the pulsed emission arises.
If the hotspot emission is modulated by the QPO, then consequently, so
are the seed photons that are reprocessed in the accretion shock. However,
the pulsed hard power-law emission is apparently not strongly modulated
by the QPO, even though the pulse fraction increases with energy. Depending
on how the modulation itself arises, two scenarios may be able to explain 
this observation. First, the QPO may be due to an oscillation in temperature
that is localized in certain region of the hotspot (e.g. its edges). In
this case the seed photons entering the shock may have a much smaller
modulation amplitude than the overall hotspot. Second, the hotspot variability
may be due to area variations, so that again the intensity of the seed
photons is largely unaffected by the QPO. 

As natural consequence of both scenarios the fractional amplitude of
pulse modulation should be lower than that of the non-pulsed QPO, as
was indeed observed.
In the case of temperature oscillations, the ratio of these amplitudes
relates to the relative area of the hotspot that is affected by the
QPO. For the area variations, on the other hand, this ratio could
indicate the size of the area variations, and hence place a limit on
the smallest and largest extent of the hotpot. How the opposite energy
dependence of the harmonic fits into this picture, however, is not
clear.

Any model that aims to explain the QPOs through a variable hotspot should 
also be able to explain the averaged shape of the pulse profile. We would
expect that variations in area and temperature each lead to a slightly different
averaged pulse profile, so that detailed modeling of the pulse profile 
shape as a function of QPO phase could be able to distinguish between the two
scenarios. Furthermore, such an approach could potentially break some the angle 
degeneracies that exists in pulse profile modeling. Such a detailed 
pulse profile modeling analysis, however, is beyond the scope of the present work.

\subsection{The 20 mHz QPO}
    The detection of a $\sim20$ mHz QPO and its harmonic has been reported for
    the \rxte observations of the 2004 outburst of IGR J00291. A similar mHz 
    QPO has also been reported for the 2008 outburst, although for those 
    observations no harmonic was detected \citep{Hartman2011}.

    By analysing the pulse amplitude modulation spectra of these data we have 
    shown that in the 2004 outburst the mHz QPOs are modulating the pulsations. 
    This result demonstrates that the methods introduced in this work
    can be used to study the coupling of weak aperiodic features with
    the periodic emission of the pulsar.

    We found that the modulation coherence spectra show a series
    of harmonic peaks with a fundamental frequency of $20$ mHz. We note
    that more harmonic peaks are visible in modulation coherence spectra
    than are visible in the direct power spectrum.
    This result suggests that pulse coherent analysis methods, such
    as discussed here, may able to extract more intricate information on 
    the driving mechanism than would otherwise be possible.

    The identification of these mHz QPOs in the context of variability observed
    in other LMXBs is uncertain. A possible interpretation is that these
    features are instances of Low-Frequency (LF) QPOs. In other LMXBs, however,
    such QPOs show a specific frequency relation with the band-limited noise
    components. As discussed by \citet{Linares2007}, the mHz QPOs in IGR J00291
    deviate from that relation by an order of magnitude. Such a deviation may
    be attributed to the anomalous morphology of the power spectrum
    \citep{Linares2007}, which has lower characteristic frequencies
    and larger fractional variability than normally observed. However,
    more recently the AMXP MAXI J0911--655 has been shown to exhibit a
    similar broad noise structure, but with LF QPOs that fall along
    the expected frequency relation \citep{Bult2017a}, casting doubt
    on this interpretation.

    From recent work on black-hole binaries a view has emerged that LF QPOs are
    due to Lense-Thirring precession of a geometrically thick inner accretion
    flow that rotates as a solid body \citep{Ingram2015,Ingram2016}. Extending
    this model to neutron stars, however, requires additional precession
    torques due to the stellar oblateness \citep{Morsink1999} and magnetic field
    \citep{Shirakawa2002} to be taken into account. The observed deviation
    from the frequency relation could then be interpreted as evidence that one of
    these additional effects is dominating the effective precession torque in IGR
    J00291. In such a model, however, the LF QPO is caused by a geometric
    effect in the accretion disk. For such a QPO to be coupled to the pulsed
    emission requires a `beating' interaction model, driven by the
    stellar magnetosphere or pulsar beam sweeping over the disk.  Because this
    interaction involves two azimuthally rotating components, it should give rise
    to a single sideband interaction. The mHz QPOs in IGR J00291, however, are
    observed in both sidebands, which strongly disfavors the precession model
    as the mechanism behind these QPOs.

    Millihertz QPOs have also been observed in a number of atoll sources
    \citep{Revnivtsev2001}. Their relation with type I X-ray bursts
    \citep{Altamirano2008c} provides strong evidence these QPOs are somehow
    related to nuclear burning on the neutron star surface, and are often
    explained in terms of marginally stable burning \citep{Heger2007}. This
    model predicts that an oscillation in the thermonuclear burning rate can
    occur in a narrow range of accretion rates. This
    interpretation poses a problem for the mHz QPOs of IGR J00291, which are
    observed throughout the outburst, and thus for a wider range of mass accretion
    rates than seen in other sources. It is worth pointing out, however, that
    there are still considerable theoretical uncertainties in the required
    conditions leading to marginally stable burning. In particular, the mass
    accretion rates used in simulations differ from those observed by an
    order of magnitude \citep{Keek2014a}.  Marginally stable burning therefore
    remain a possible explanation for the mHz QPOs observed in this source.

    Variations in the mass accretion rate offer another class of
    models that might be responsible for QPOs that couple to the
    pulsations. In this scenario the variable accretion rate is
    directly responsible for an oscillation in stellar hotspot
    emission.  In the specific case of IGR J00291,
    however, an issue is that there are few mechanism that have
    sufficiently long timescales to produce mHz variability
    \citep[see][for a detailed discussion]{Ferrigno2017}. Our
    detection of QPO/pulse coupling does not provide further
    insight. Detailed analysis of the energy spectra of the direct and
    modulation QPOs may be able to place additional constraints on such
    models, however, given limited data quality this was not possible
    with the available \rxte observations.

\subsection{Comparing the QPOs}
    Comparing the coupling characteristics for the \xmm mHz QPO with 
    the \rxte mHz QPOs gives some clues on whether these phenomena
    share a common physical mechanism. For one, we note that both QPOs
    couple to the pulsations, and both appear in symmetrically in
    the positive and negative sidebands. This alone is already a strong
    indication that these features are somehow related to emission originating
    on the stellar surface, and most likely the stellar hotspot.
    
    We also observed a tentative $6$ mHz QPO modulating the pulsations
    in the 2004 outburst, which is not directly visible in the regular
    power spectrum.
    If this component is indeed real, it would suggest that whatever mechanism
    is causing the very soft $8$ mHz QPO in the \xmm data, is also active in
    the 2004 outburst, but at a slightly lower frequency. 
    This interpretation does not imply a relation between 
    the $8$ mHz QPO and the $20$ mHz QPOs, however, with the striking resemblance
    between the modulation coherence spectra, especially between the harder
    \xmm band and the \rxte observations, 
    a single, non-trivial oscillation of the stellar hotspot seems 
    the most plausible avenue toward explaining all these millihertz features
    and the characteristics of their coupling with the pulsed emission.

\acknowledgments
We would like to thank the referee for detailed comments that helped improve
the presentation of this work.
PB was supported by an NPP fellowship at NASA Goddard Space Flight 
Center. MvD and MvdK acknowledge support from the Netherlands Organisation 
for Scientific Research (NWO).

\bibliographystyle{fancyapj}

\end{document}